\documentclass[twocolumn,prl,sort&compress,floatfix]{revtex4-1}

\usepackage{amssymb}
\usepackage{amsmath}
\usepackage{mathrsfs}
\usepackage{bm}
\usepackage[dvipsnames]{xcolor}
\usepackage[sort&compress]{natbib}

\usepackage{graphicx}

\usepackage{color}
\definecolor{red}{rgb}{1,0,0}
\definecolor{blue}{rgb}{0,0,1}

\newcommand{\dmi}[1]{{\color{Black} #1}}

\makeatletter

\let\c@table\c@figure
\makeatother

\begin{document}

\title{Maximally Stiffening Composites Require Maximally Coupled \\ Rather Than Maximally Entangled Polymer Species}

\author{Davide Michieletto$^{a,b}$, Robert Fitzpatrick$^c$, Rae M Robertson-Anderson$^c$}
\affiliation{$^a$School of Physics and Astronomy, University of Edinburgh, Edinburgh, UK}
\affiliation{$^b$ MRC Human Genetics Unit, Institute of Genetics and Molecular Medicine, University of Edinburgh, Edinburgh EH4 2XU, UK}
\affiliation{$^c$Department of Physics and Biophysics, University of San Diego, San Diego, CA, USA}

\begin{abstract}
	\textbf{
    \centering ABSTRACT\\
    \flushleft
	Polymer composites are ideal candidates for next generation biomimetic soft materials because of their exquisite bottom-up designability. However, the richness of behaviours comes at a price: the need for precise and extensive characterisation of material properties over a \dmi{highly-dimensional} parameter space, as well as a quantitative understanding of the physical principles underlying desirable features. Here we couple large-scale Molecular Dynamics simulations with optical tweezers microrheology to characterise the viscoelastic response of DNA-actin composites. \dmi{We discover that the previously observed non-monotonic stress-stiffening of these composites is robust, yet tunable, in a broad range of the parameter space that spans two orders of magnitude in DNA length. Importantly, we discover that the most pronounced stiffening is achieved when the species are maximally coupled, i.e. have similar number of entanglements, and not when the number of entanglements per DNA chain is largest. We further report novel dynamical oscillations of the microstructure of the composites, alternating between mixed and bundled phases, opening the door to future investigations.} The generic nature of our system renders our results applicable to the behaviour of a broad class of polymer composites. 
	} 
\end{abstract}

\maketitle

\section{Introduction}
Composite systems of polymer species with distinct physical and/or chemical properties are emerging as promising candidates for next generation multifunctional materials due to the ease and breadth in which their material properties can be finely tuned by the properties of the constituents~\cite{Lavine2018,Eder2018}. \dmi{Composites or polymer mixtures have been shown to display emergent mechanical and rheological properties that are not a simple superposition of the corresponding single-component systems but strongly depend on, e.g., entropic or enthalpic interaction between the species~\cite{Stiakakis2005,Truzzolillo2011}, their softness~\cite{Mayer2008}, or topology~\cite{Kapnistos2008}}. 
In fact, these behaviours are often completely unexpected given the starting materials~\cite{Burla2019,Du2019,Jaspers2017} and include nonlinear stress-stiffening~\cite{Storm2005,Lin2011,Ricketts2018}, \dmi{asymmetric caging in soft colloidal glasses~\cite{Mayer2008}}, negative normal stress~\cite{Janmey2007,Pelletier2009}, and strength with simultaneous lack of brittleness~\cite{Eder2018}. While we have only recently begun to appreciate the physics underlying the emergent material properties of polymer composites, nature has already exploited their design principles to allow for multifunctional mechanics and dynamic processes in the cell nucleus~\cite{Baum2014,Hameed2012,Marko2008}, cytoskeleton~\cite{Broedersz2008,Brangwynne2008,Jaspers2017} and extracellular matrix~\cite{Kasza2007,Jaspers2017}. As such, while understanding the design principles of polymer composites is important for synthetic applications~\cite{Lee2005,Lee2002}, it also directly informs complex biological networks and paves the way for new biomimetic materials~\cite{Eder2018}.

Inspired by all this, we designed a suite of entangled solutions of flexible DNA (persistence length $l_p \simeq$ 50 nm) and semiflexible actin filaments ($l_p \simeq$ 10 $\mu$m) with physical properties that can be tuned over a wide parameter space. These two ubiquitous biopolymers have been separately studied extensively as model flexible and semiflexible polymer systems to elucidate questions in polymer physics and engineering  ~\cite{Teixeira2007,Gardel2004,Liu2006}. These studies have shed important light on our understanding of entangled polymers and the tube models used to describe them~\cite{Isambert1996,Robertson2007,Robertson2007a,Wang2010}. Yet, there is limited knowledge on how composites of these two iconic biopolymers behave~\cite{Lai2008a,Negishi2010,Fitzpatrick2018}. Here, we \dmi{tackle the vast multi-dimensional parameter space of our previously introduced custom-engineered actin-DNA composites to elucidate the role that DNA length, number of entanglements, and dynamics play in their emergent rheological properties.} Our combined numerical and experimental approach - coupling large-scale Brownian Dynamics simulations with optical tweezers microrheology measurements - \dmi{is uniquely positioned to characterise the rich physics underlying the mechanics of biopolymer composites and it serves as proof of principle for precision design of next generation polymer composites}.

We choose to fix the total polymer concentration of our composites to a unique value ($c=0.8$ mg/ml) in which the entanglement timescales and tube diameters of both single-component systems (DNA or actin) are matched, such that the principle lengthscales and timescales of the system remain fixed as we vary the relative concentrations or lengths of the two components. 
We explore a wide parameter space of composite design by independently varying the the DNA contour length $L_{DNA}$ over 2 orders of magnitude and the mass fraction of actin $\phi_A = c_{actin}/c$ from 0 to 1. We map how the rheological properties of the resulting composites are tuned by these design parameters, and our simulations reveal the macromolecular interactions that govern the observed emergent properties.

\begin{figure*}[t!]
	\centering
	\includegraphics[width=1\textwidth]{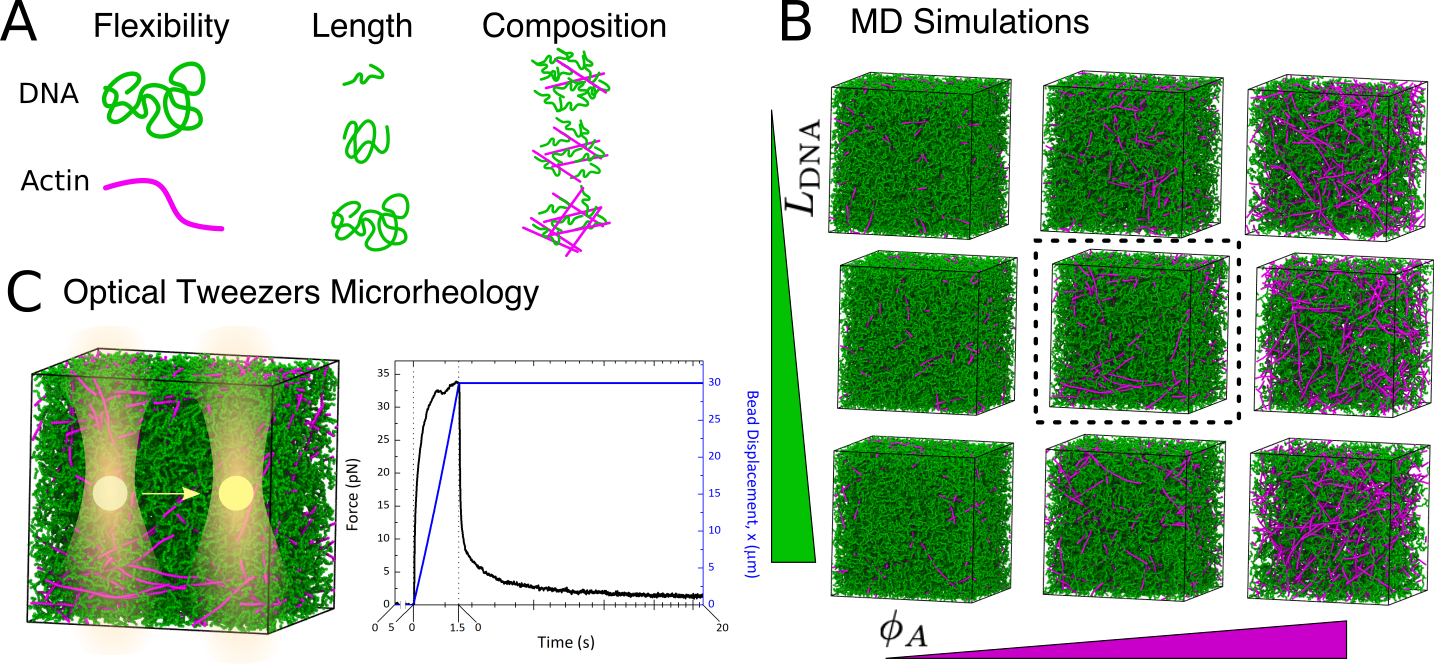}
	\caption{\textbf{Comprehensive approach to the design and characterization of polymer composites.}  \textbf{A} We design polymer composites using varying mass ratios of polymers with different flexibilities and lengths. \textbf{B} Snapshots from BD simulations of an array of DNA-actin composites, exploring a 2-dimensional parameter space in which  $L_{DNA}$ and $\phi_A (=c_{actin}/c)$ are varied. We show representative images for $\phi_A=0.25,0.5,0.75$ and $L_{DNA} = 3.67, \, 39, \, 96$ $\mu$m. \textbf{C} For each of the values of $L_{DNA}$ and $\phi_A$ shown in {B}, we experimentally realise the composite and perform nonlinear optical tweezers microrheology measurements. In these experiments, an optically trapped microsphere (4.5-$\mu$m diameter) embedded in the composite is displaced 30 $\mu$m (blue) at 20 $\mu$m/s. The force (black) is measured before (equilibrium), during (strain) and after (relaxation) the bead displacement. 
    }
	\label{fig:panel1}
	\vspace*{-0.0 cm}
\end{figure*}

\dmi{Importantly, we recently discovered that these composites exhibit unique viscoelastic properties that depend non-monotonically on the actin fraction $\phi_A$~\cite{Fitzpatrick2018}. Yet, how their mechanical behaviour depends on the myriad of other tunable system parameters, such as polymer length and number of entanglements, remains unknown.}   
\dmi{In particular, one may expect that increasing the length of the flexible species -- and thus the total number of entanglements in the system -- would undoubtedly increase the elastic response of the composite. However, here we report computational and experimental evidence rebutting this naive expectation.}

\dmi{ 
Both our simulations and experiments unambiguously demonstrate that the previously discovered stress stiffening is surprisingly most pronounced in composites in which the number of entanglements per chain is comparable between species}. \dmi{They further show that the existence of an optimum actin fraction for enhanced stress-stiffening is broadly robust but can be tuned by changing the length of the flexible species}.
\dmi{Finally, our large-scale simulations allow us to detect, for the first time, intriguing dynamical oscillations in the micro-structure of the composites which are likely important to understanding potential heterogeneities in the mechanical properties, or even failure~\cite{Aime2018}, of polymer composites.} 

Since our actin-DNA system is a generic model for any polymer network in which stiff and flexible components interact, our work uncovers the generic physics underlying the mechanical response of this broad class of materials - ubiquitous in biology as well as industry - and contributes to revealing their universal design principles. More generally, we demonstrate a strategy to systematically and extensively characterise the parameter space of polymer composites, and explore in full their rich behaviours, through a combination of large-scale Brownian Dynamics simulations and optical tweezers experiments (see Fig.~\ref{fig:panel1}). This strategy can be applied to a wide range of polymer composites and biomimetic networks, and can therefore fast-track the discovery of desirable material properties in polymeric materials.

\section{Results}

\subsection{Polymer Composite Design and Physical Parameters}

As described in the Introduction, we choose to fix the overall concentration of our composites to $c=0.8$ mg/ml at which the entanglement tube diameter $a$ and entanglement time $\tau_e$ for solutions of only DNA or actin are matched~\cite{Fitzpatrick2018}. This allows us to tune the relative concentrations of flexible polymers (DNA, $l_p \simeq 50$ nm) and semiflexible ones (actin, $l_p\simeq 10 \, \mu$m), as well as the length of the flexible component, while preserving key polymer physics quantities. Moreover, this concentration is above the entanglement concentration for both polymers but below the concentration in which actin displays nematic ordering~\cite{Gurmessa2016,Kaz1996,Lai2008a}. We chose to use DNA lengths of $L_{DNA} =3.67, \, 38, \, 96$ $\mu$m which correspond to primitive path lengths, or tube lengths, of  $L_{0, DNA} \simeq 0.5, \, 5, \, 12$ $\mu$m~\cite{Gurmessa2016,Chapman2014prl}. For semiflexible actin filaments $L_{0,actin} \simeq L_{actin}$ and in all composites $L_{actin} \simeq 7 \mu$m independent of $\phi_A$ and $L_{DNA}$. Thus, $L_{0, DNA}$ spans from $\sim 10\times$ below to $\sim 2\times$ above $L_{0,actin}$. The number of entanglements per DNA chain $Z_{DNA}$ likewise scales with $L_{0, DNA}$.

The key lengthscale governing dynamics of entangled polymers is the tube diameter $a$. For flexible chains the tube diameter is $a_{flex}=(2 l_p l_e^{flex})^{1/2}$, where $l_e^{flex}={4 c R T}/ {5 G_0 M_{bp}}$ is the polymer length between entanglements, $G_0$ is the plateau modulus ~\cite{Doi1988,Chapman2014prl}, and $M_{bp}=650 g/mol$. Using experimental values for $G_0$~\cite{Teixeira2007,Chapman2014a} we compute the tube diameter for DNA in our composites as $a_{flex}\simeq 0.64$ $\mu$m.
For semiflexible polymers such as actin  $a_{semiflex} \simeq  l_e^{semiflex} = \xi^{4/5} l_p^{-1/5} $, where $\xi=0.3/\sqrt{c}$ is the mesh size~\cite{Gardel2003,Lang2018,Isambert1996,Gurmessa2016}. Using this expression we compute $a_{semiflex} \simeq 0.66 \, \mu \text{m}$ for actin in our composites, purposefully nearly matched to the tube diameter for DNA.
\dmi{It should be noted that since $l_p$ and $c$ are fixed the two entanglement lengths remain constant for all values of $\phi_A$}.

The key timescale governing the onset of entanglement effects in polymer systems is the entanglement time $\tau_e$, defined as the time required to relax one entanglement length of the polymer~\cite{Doi1988}. For flexible chains $\tau_e^{flex} = {a^4}/{[24 R_g D]}$ where $R_g$ and $D$ are the radius of gyration and diffusion coefficients in dilute conditions (experimentally computed for DNA in Ref.~\cite{Robertson2006}). Because $R_g D$ is independent of polymer length, for all $L_{DNA}$ considered here $\tau_e^{flex}\simeq 0.04 \, s$. Actin filaments in our system have a similar value of $\tau_e^{semiflex} = \beta \zeta \xi^{16/5} l_p^{-1/5} \simeq 0.05 \, s$ where $\beta=1/k_BT$ and $\zeta$ is the friction coefficient~\cite{Isambert1996,Kaz1996}.
 
 While $a$ and $\tau_e$ are independent of DNA length, the key relevant parameter that can be tuned by $L_{DNA}$ is the number of entanglements per DNA chain $Z_{DNA}=L_{0,DNA}/a_{flex}$, which takes values of $\sim$1, 10 and 22 for the three chosen DNA lengths. Because $Z_{actin} \simeq 10$ for all composites \dmi{-- as $l_e^{semiflex}$ depends only on $l_p$ and $c$ which are fixed --} the entanglement ratio $ZR = Z_{DNA}/Z_{actin}$ is $\sim$0.1, 1, and 2.3 for the three DNA lengths used here. Note that the composites in which $L_{DNA}=39 $ $\mu$m are the ones in which the number of entanglements per chain is the same for both polymer species, i.e. $Z_{DNA} \simeq Z_{actin}$ or $ZR$ $\simeq 1$. 

\begin{figure*}[t]
	\centering
	\includegraphics[width=0.95\textwidth]{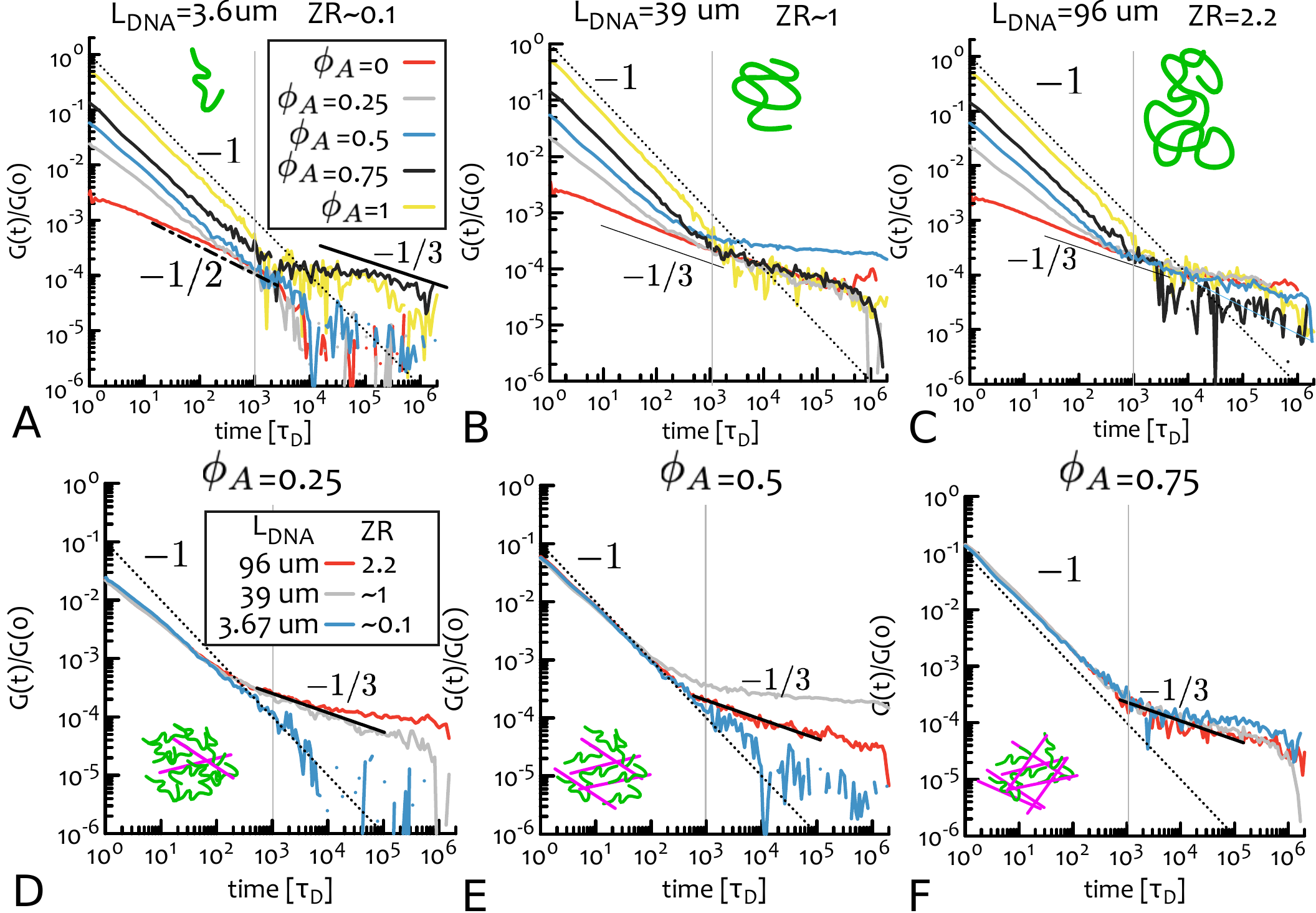}
	\vspace{-0.4 cm}
	\caption{\textbf{Stress relaxation of polymer composites reveals emergent elastic plateaus.} \textbf{A-C} The stress relaxation modulus from BD simulations, normalised by its $t=0$ value $G(t)/G(0)$ is shown for composites with varying $\phi_A$ values as indicated in the legend and $ZR$ values of $\sim0.1$ (\textbf{A}),  $1$ (\textbf{B}), and  $2.2$ (\textbf{C}). \textbf{D-F} The same data shown in \textbf{A-C} plotted for different $ZR$ values and fixed $\phi_A$ values of 0.25 (\textbf{D}), 0.5 (\textbf{E}) and 0.75 (\textbf{F}). For all systems, the transition between different power law regimes occurs at ~$10^3 \tau_D \simeq 0.04s$ (grey vertical line), which corresponds to $\tau_e$ for both polymer species. While single-component systems (i.e. $\phi_A$ = 0, 1) display no rubbery plateaus, composite systems with certain ($ZR$, $\phi_A$) combinations exhibit strong plateaus-- most notably ($\sim$1, 0.5), ($\sim$0.1, 0.75),  and ($\sim$2.2, 0.25). Unexpectedly, composites with the longest DNA (and thus largest number of entanglements per chain) display the weakest elastic plateaus. Dashed and dotted lines are guides for the eye and mark different power-law behaviours.
    }
    	\vspace{-0.2 cm}
	\label{fig:panel1_gt}	
\end{figure*}
\subsection{Stress Relaxation of Polymer Composites}
In order to predict the mechanical and structural properties of the composites, we perform large-scale Brownian Dynamics (BD) simulations of a coarse-grained model. Specifically, we consider a mix of bead-spring chains in which each bead represents a segment of DNA or actin. We choose a coarse-grain level at which each polymer bead is $25$ nm, equivalent to half the DNA persistence length and 9.2 actin monomers (see Materials and Methods). Similar coarse-grained models for DNA and actin have been employed in the literature~\cite{Lang2018,Fitzpatrick2018} and have been shown to well capture the dynamics of experimental systems. 

From the simulations, we first extract the stress relaxation modulus $G(t)$~\cite{Doi1988}, which we obtain from the stress tensor $\sigma^{ab}$ defined as
\begin{equation}
 V \sigma^{ab} \equiv \sum_{k} m_k v_k^a v_k^b + \dfrac{1}{2}\sum_{i}\sum_j F_{ij}^a r_{ij}^b \, ,
\end{equation}
where $a,b$ are the Cartesian components, $m$ is the mass of the bead, $v$ its velocity, $F$ the force between beads, $r$ their distance, $V$ the volume of the system, and $k,i,j$ running over all beads in the system. The stress tensor is pre-averaged for $t_a=100$ integration steps , i.e. $\bar{\sigma}^{ab}(t) = \sum_{dt=-t_a/2 +1 }^{t_a/2} \sigma^{ab}(t+dt)$, and the auto-correlation function computed on the fly using a multi-tau algorithm~\cite{Ramirez2010}, finally giving the stress auto-correlation as $G^{ab}(t)=V \langle \bar{\sigma}^{ab}(t) \bar{\sigma}^{ab}(0) \rangle/k_BT$. The stress relaxation modulus is then obtained by averaging its out-of-diagonal components, i.e. $G(t) \equiv (G^{xy}+G^{yz}+G^{xz})/3$. 

Fig.~\ref{fig:panel1_gt} displays $G(t)$ for different choices of $\phi_A$ and $L_{DNA}$. All composites exhibit two phases of power-law relaxation with a crossover in scaling regimes occurring at time $10^3 \tau_D  \simeq 0.04$ s independent of $\phi_A$ and $L_{DNA}$ \dmi{($\tau_D \equiv \sigma^2/D$ is the definition of a Brownian time and the time-unit of our simulations, $\sigma$ is the size of a bead and $D$ its diffusion constant)}. This timescale, which appears to play a critical role in dynamics, is remarkably close to the entanglement time of both polymer species ($\tau_e^{semiflex} \simeq \tau_e^{flex} \simeq 0.04$ s), suggesting that entanglements play a key role in the emergent behavior that we observe. We thus conjecture that varying the number of entanglements per chain and thus the mutual entanglements of the two polymer species (by varying $L_{DNA}$ and thus $ZR$) may play an important role in the mechanics. We discuss this phenomenon further below.  

For pure  solutions of short DNA  ($\phi_A=0$, $L_{DNA}= 3.67 \, \mu$m, $Z_{DNA} \simeq 1$) we observe $ G(t) \sim t^{-1/2}$ , in agreement with the Rouse behaviour of poorly entangled chains~\cite{Rubinsteina}, whereas we find $G(t)\sim t^{-1/3}$ for longer DNAs, in agreement with previously reported values for entangled DNA~\cite{Fitzpatrick2018,Teixeira2007}. Conversely, pure actin networks ($\phi_A=1$) display an initial decay of $G(t)\sim $  $t^{-1}$, in between $t^{-2/3}$ and $t^{-5/4}$ predicted in Ref.~\cite{Pasquali2001}. Collectively, these results demonstrate the validity of our simulations in capturing the dynamics of real DNA and actin systems.

Upon mixing DNA and actin ($0 < \phi_A < 1$), we observe emergent mechanics. Composites with all three $L_{DNA}$ display elastic plateaus for $t > 0.04~s$ that are not seen in either single-component network ($\phi_A=0$ or $1$). This signature of elastic behaviour and stiffness is most apparent for the composite with $ZR$ $\simeq$ 1 and $\phi_A=0.5$ (Fig.~\ref{fig:panel1_gt}B). There is also a notable plateau for $ZR \simeq 0.1$ and $\phi_A =0.75$ and a weaker one for $ZR \simeq$ 2.2 and $\phi_A =0.25$ (Fig.~\ref{fig:panel1_gt}A,C). The dependence of the elastic response on  $L_{DNA}$ can be seen more clearly in Fig.~\ref{fig:panel1_gt}D-F) which shows that as $\phi_A$ increases the DNA length required for composites to exhibit the strongest plateau decreases. Further,  the composites that display the weakest $\phi_A$-dependent plateaus are surprisingly those with the longest DNA chains and thus the largest number of entanglements per chain ($Z_{DNA} \simeq 22$). \dmi{This is a priori unexpected, as one may argue that longer DNA molecules increase the total entanglements in the system and hence should yield larger network stiffening. }

\begin{figure}[t!]
	\centering 
	\includegraphics[width=0.45\textwidth]{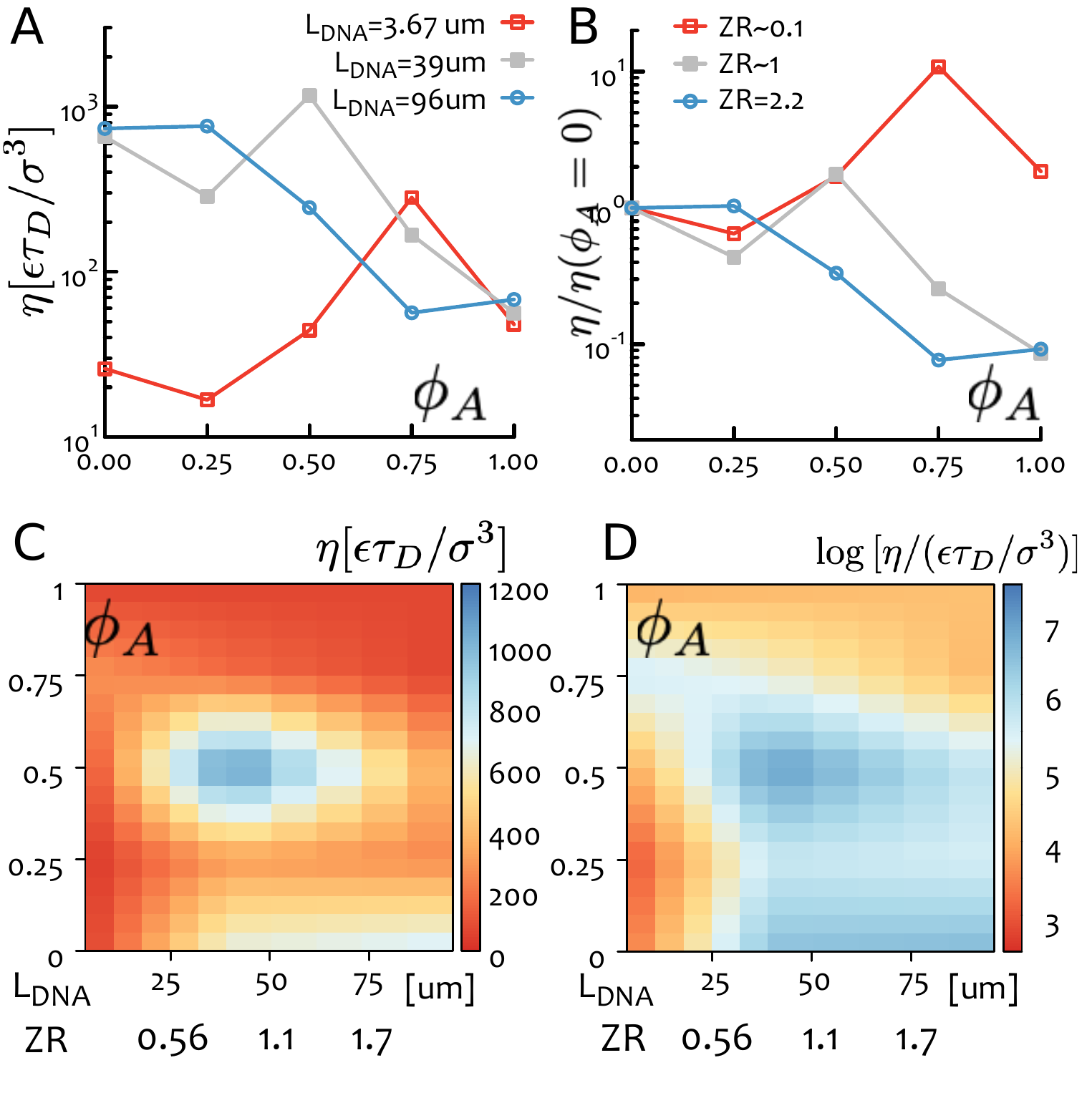}
	\vspace{-0.5 cm}
	\caption{\textbf{Zero-shear viscosity exhibits a non-monotonic dependence on mass composition}. \textbf{A} Zero-shear viscosity $\eta$ (see eq.~\eqref{eq:visc}) as a function of $\phi_A$ for varying $ZR$ (and $L_{DNA}$) values shown in legend. \textbf{B} Zero-shear viscosity shown in \textbf{A} normalized by the corresponding $\phi_A$=0 value, $\eta/\eta(\phi_A=0)$. \textbf{C-D} Heat map of $\eta$ over the full parameter space determined via numerical interpolation of our 5x3 ($ZR,\phi_A$) data points shown in \textbf{A, B} and plotted on linear (\textbf{C}) and log (\textbf{D}) scales. \textbf{C} shows a clear maximum at ($ZR$,$\phi_A)=(\sim$1, 0.5), and \textbf{D} highlights the inverse relationship between $ZR$ and $\phi_A$, namely as $ZR$ increases the optimum $\phi_A$ at which the composites display the largest viscosity decreases.}
		\vspace{-0.48 cm}
	\label{fig:panel1_eta}	
\end{figure}

\begin{figure}[b!]
	\centering
	\includegraphics[width=0.45\textwidth]{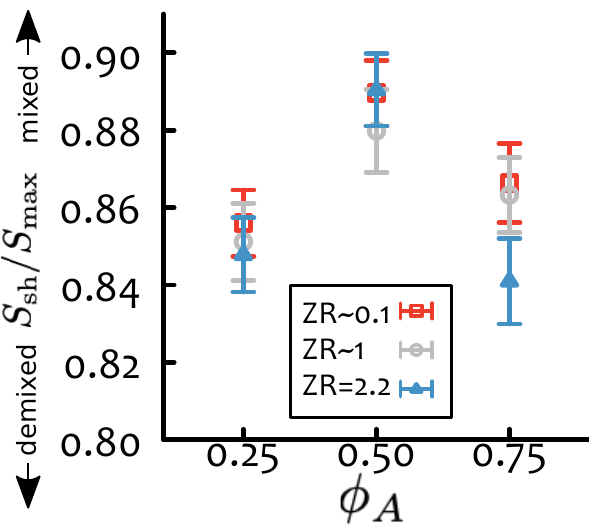}
	\caption{\textbf{De-mixing of polymer composites, captured by the Shannon Entropy, depends non-monotonically on $\phi_A$.}  Shannon entropy $S_{sh}$ averaged over the whole simulation time and divided by the corresponding maximum attainable value $S_{max}$ for each ($ZR$, $\phi_A$). As shown, the maximum entropy, corresponding to the least amount of de-mixing, is achieved at $\phi_A=0.5$ for all $ZR$ values. }
	\label{fig:entropy_mixing_tot}	
\end{figure}
From $G(t)$ we can extract the zero-shear viscosity~\cite{Doi1988,Halverson2011dynamics}, i.e. 
\begin{equation}
\eta = \int_0^\infty G(t)dt \, ,
\label{eq:visc}
\end{equation} 
which is a measure of the resistance of each composite to infinitely small shearing. As shown in Fig.~\ref{fig:panel1_eta}, $\eta$ readily reflects a non-monotonic dependence on $\phi_A$, as well as sensitivity to $Z_{DNA}$. As with the elastic plateau in $G(t)$, the zero-shear viscosity attains its global maximum for ($ZR, \phi_A) = (\sim$1, 0.5), and displays other maxima for ($ZR, \phi_A) = (\sim0.1, 0.75$) and $(ZR, \phi_A) = (\sim2.2, 0.25)$. 
Further, in line with the behaviour of $G(t)$, the non-monotonic $\phi_A$-dependence is weakest for the longest DNA, with the maximum value reached close to that for $\phi_A=0$.  

To fully map the phase space of composite elasticity and stiffness, we linearly interpolate the 5x3 simulated data points and plot as a heat-map (Fig.~\ref{fig:panel1_eta}C-D). As shown, one can clearly notice a marked maximum in viscosity when actin and DNA are perfectly balanced, i.e. $Z_{DNA} \simeq Z_{actin}$ and $\phi_{DNA}=\phi_{actin}$. Fig.~\ref{fig:panel1_eta}D also highlights that, as $L_{DNA}$ increases, the enhanced stiffening behaviour shifts to lower $\phi_A$ values (as also seen in Fig.~\ref{fig:panel1_gt}C-F). 

\dmi{These results  not only confirm the important role that synergistic interactions play in the elasticity of the biopolymer composites~\cite{Fitzpatrick2018}, but also demonstrate that composites can be designed to display a wide range of mechanical properties by independently tuning the length of the flexible species (i.e. $L_{DNA}$) and mass composition (i.e. $\phi_A$). In particular, we discover that the optimum mass composition to achieve maximum stiffening can be adjusted via the length of the flexible species. At the same time, the strength of the response has an unexpected dependence on the number of entanglements as it displays a maximum when the number of entanglements of the two species is comparable rather than largest.}


\subsection{Entropy of Mixing Quantifies Composite Macrostructure}
To shed light on the mechanical properties described above we turn to characterizing the macro- and microscopic arrangements of DNA and actin polymers comprising the composites. 

The de-mixing, or phase separation, of a system made of $i=1, \dots, M$ types of monomers can be quantified by computing the Shannon entropy (i.e. entropy of mixing)~\cite{Brandani2013}
\begin{equation}
S(b) \equiv - \sum_{i=1}^M \rho_{i}(b) \log{\rho_{i}(b)}  \, .
\end{equation}
In the previous equation, we tile the system with $N_b$ boxes and define $\rho_{i}(b)=n_i/\sum_j^M n_j(b)$ as the mole (or number) fraction of particles of type $i$ in box $b$. The Shannon entropy is the average of $S(b)$ over all the boxes tiling the system, i.e. 
\begin{equation}
S_{sh}\equiv \dfrac{\sum_{b=1}^{N_b} S(b)}{N_b} \,.
\label{eq:entropy} 
\end{equation}
The maximum value $S_{max}$ is attained when the system is fully mixed, i.e.  $S_{max} \equiv - \sum_i^M \rho_{i} \log{\rho_{i}}$, where $\rho_i$ is the overall fraction of monomers of type $i$ in the system. For instance, in the case of a symmetric binary fluid, $S_{max}=-\log{0.5} \simeq 0.69$; instead, in this work we consider two types of monomers -- those belonging to DNA or actin -- at varying relative concentrations, and we thus have a composition dependent maximum entropy, i.e. $S_{max}=S_{max}(\phi_A, L_{DNA})$. 

Deviations from the fully mixed state can thus be reported and compared across conditions via the ratio $S_{sh}/S_{max}$, as shown in Fig.~\ref{fig:entropy_mixing_tot} where we plot the time-average of $S_{sh}/S_{max}$ for different values of $\phi_A$ and $ZR$. 
As shown, deviations from the fully mixed state are typically of the order to 10-20\%. In other words, while some de-mixing occurs within the system, the two polymer species are still largely interacting with one another, in contrast to the macro-scale phase separation reported in other actin-DNA systems in which much higher concentrations and/or micro-scale confinement are used~\cite{Lai2008a,Negishi2010}. Another surprising feature is that the minimum deviation from $S_{max}$ is found for $\phi_A=0.5$, which is the same actin fraction that exhibits the most extreme elastic response. While emergent behavior in composite systems is often attributed to large-scale phase separation, these notable results indicate that it is the mixing of the two species that enables the most pronounced emergent behavior. 
\dmi{It should be noted that we expect these results to qualitatively hold even if choosing different sizes of boxes and ways to tile our system (in this case we chose to use 512 boxes for a system of size $L=100 \sigma$).}

\begin{figure*}[t!]
	\centering
	\includegraphics[width=0.95\textwidth]{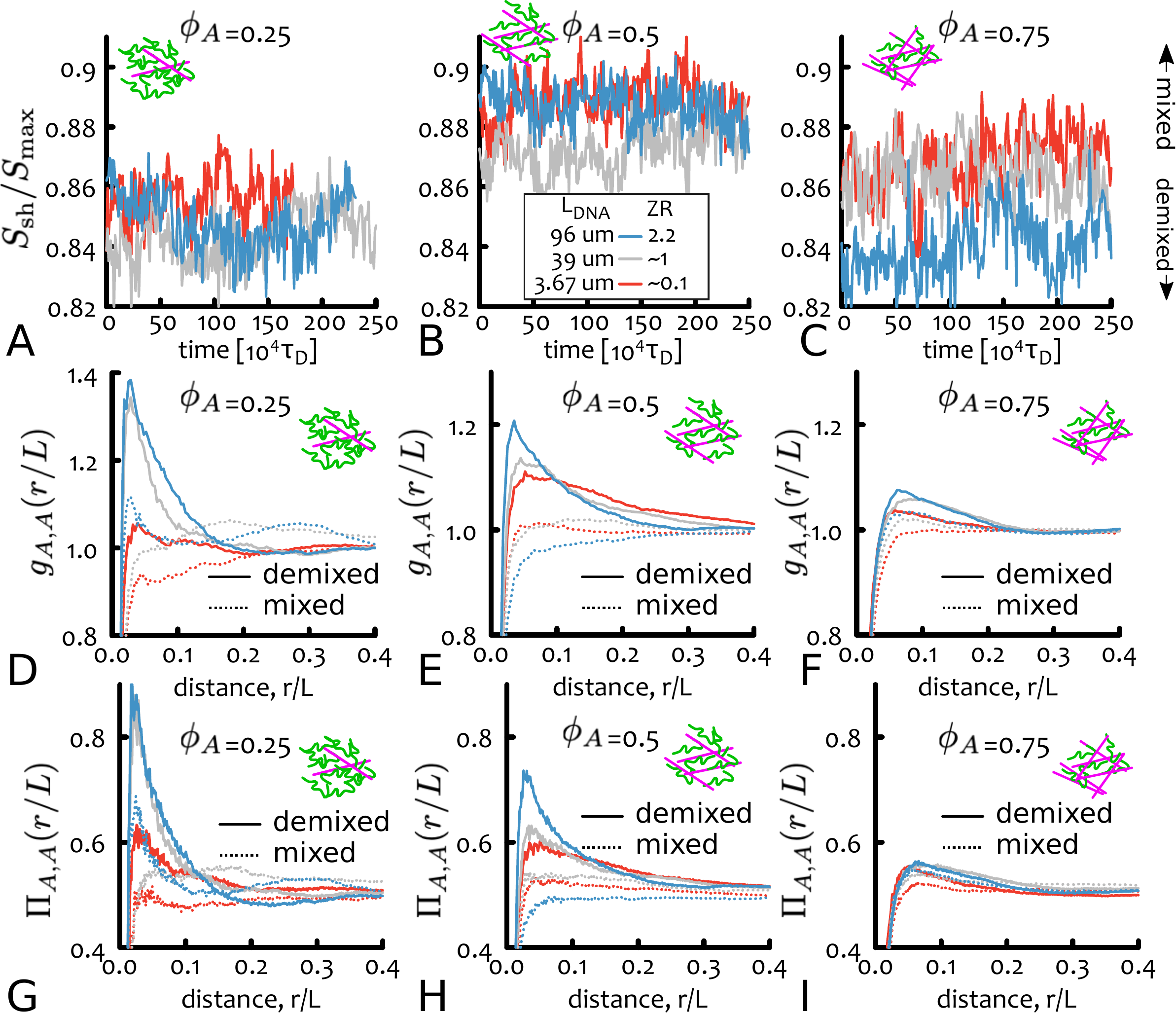}
	\caption{\textbf{ Actin polymers in composites fluctuate between bundled (de-mixed) states and unbundled (well-mixed) states.} \textbf{A}-\textbf{C} The Shannon Entropy $S_{sh}/S_{max}$ for all ($ZR$, $\phi_A$) values show significant temporal fluctuations which correspond to dynamic structural rearrangements between states in which DNA and actin are well-mixed and ones in which they are partially de-mixed. Maxima in $S_{sh}(t)/S_{max}$ curves correspond to mixed states whereas minima correspond to de-mixed states.  \textbf{D}-\textbf{F} The actin-actin radial distribution function $g_{A,A}(r/L)$ versus distance $r/L$, computed for both maximum and minimum $S_{Sh}/S_{max}$ values, show that actin fluctuates between being uniformly distributed throughout the composite (mixed, monotonic rise to unity) and forming short-range aggregates (de-mixed, peaks at short lengths). \textbf{G}-\textbf{I} The actin-actin nematic correlation function $\Pi_{A,A}(r/L)$ for each ($ZR$, $\phi_A$) exhibit similar trends as $g_{A,A}(r)$, showing that the actin aggregation is in fact nematic bundling of actin filaments. } 
	\label{fig:entropy_mixing}
\end{figure*}

Finally, we highlight that the instantaneous, i.e. time-dependent, Shannon entropy for the composites displays fluctuations in time which are indicative of dynamical rearrangements of the composites micro-structure (Fig.~\ref{fig:entropy_mixing}A-C). 
\dmi{These small-scale reorganizations have not been previously reported in polymer composites as they elude the spatiotemporal resolution that can be achieved with current experimental approaches. As such, we characterise this novel phenomenon in more detail in the next section.}



\subsection{Transient Aggregation and Nematic Bundling Characterise the Composite Microstructure}

The micro-scale organisation of the composite can be quantified by computing the radial distribution function (RDF) for monomers belonging to DNA and/or actin polymers, i.e. 
\begin{equation}
rdf_{x,y}(r) = \langle \delta( |\bm{r}_{x,i} - \bm{r}_{y,j} | - r) \rangle \, ,
\end{equation}
where $\bm{r}_{x,i}$ is the position of bead $i$ belonging to polymer species $x$ (with $x=\{A,D\}$ for either actin or DNA) and the average is taken over time and monomers. This quantity measures the probability of observing two beads of types $x$ and $y$ to be found at distance $r$ from each other. We normalise with respect to the expected uniform behaviour to get 
\begin{equation}
g_{x,y}(r) = \dfrac{rdf_{x,y}(r)}{4 \pi \rho_y r^2 dr}\, ,  \label{eq:rdf}
\end{equation}
where $dr$ is the thickness of the spherical shell used to partition the region into bins and $\rho_y$ is the monomer density of species $y$. 
For a purely uniform and random distribution of monomers, $g_{x,y}(r)$ increases monotonically from 0 to 1 over a lengthscale dictated by the average spacing between monomers. On the contrary, aggregation causes the RDF to display a peak at short lengthscales. 

A similar quantity that is useful in the following is the nematic correlation function (NCF) which we here define as
\begin{align}
\Pi_{A,A}(r) &= \langle | \bm{t}_i \cdot \bm{t}_j | \rangle_{g(r)} =  \\
&= \dfrac{ \sum_I \sum_{J>I} \sum_{i \in I} \sum_{j \in J} | \bm{t}_i \cdot \bm{t}_j | \delta(|\bm{r}_i-\bm{r}_j| - r)}{ \langle  \delta(|\bm{r}_{i}-\bm{r}_{j}| - r) \rangle } \notag \, .
 \label{eq:nematic}
\end{align} 
In this equation, the indexes $I,J$ run over actin polymers while $i,j$ run over the beads on the polymers and $\bm{t}_{i}$ denotes the tangent to bead $i$ along the backbone. The absolute value is taken to record nematic ordering, i.e. irrespective of orientation of the tangents; it is zero if the segments are orthogonal and unity if parallel. The value of the dot product is recorded to occur at distance $r$ between the two beads $i$ and $j$. Finally, we normalise by the radial distribution function to account for local variations in density of monomers. In the isotropic case, all the pairs display random orientation and hence the equation above is equivalent to averaging over numbers uniformly distributed between 0 and 1, i.e. $\Pi(r)=1/2$. Values $\Pi_{A,A}(r)>1/2$ at short distances suggest nematic alignment and bundling of actin filaments.

At this stage we should recall that $S_{sh}/S_{max}$ fluctuates in time (Fig.~\ref{fig:entropy_mixing}A-C). Thus, we refrain from computing RDF and NCF curves averaged over the whole simulation and instead calculate eq.~\eqref{eq:rdf} within short time windows ($20 \, 10^4 \tau_{B}$ time steps) that are long enough to give us good averages but short enough to characterise transient structures. Each of these curves can be mapped to a short time-window in the trajectory of $S_{sh}(t)/S_{max}$ (Fig.~\ref{fig:entropy_mixing}A-C) and can therefore inform us on transient micro-scale structures assumed by the polymer species.  

In Fig.~\ref{fig:entropy_mixing}D-F we show RDF and NCF curves for actin-actin monomers, i.e. $g_{A,A}(r/L)$ and $\Pi_{A,A}(r/L)$.
We identify two main behaviours: (i) RDFs and NCFs with peaks at short distances, which correspond to minima of $S_{sh}/S_{max}$ and thus indicate partially demixed phases with actin bundling; and (ii) RDFs and NCFs curves with weak or absent peaks, corresponding to maxima of the Shannon entropy and thus largely mixed phases.  These micro-scale structures alternate in time suggesting dynamic arrangements of the polymer species within the composite. On the other hand, DNA-DNA and actin-DNA RDFs, i.e. $g_{D,D}(r/L)$ and $g_{D,A}(r/L)$, display a monotonic increase to unity for all composites and time-windows, indicating that (iii) DNA is uniformly distributed throughout the composite and (iv) DNA and actin polymers are co-entangled and interacting. 

Moreover, the degree of actin bundling (i.e. the strength of the peaks and the lengthscale over which they persist) depends on both $\phi_A$ and $ZR$. We find that the bundling is strongest (largest peaks) for the longest DNA for all $\phi_A$ values, and decreases with decreasing $L_{DNA}$. This short-scale aggregation, which leads to an increase of the stiffness of actin fibers (comprised of several filaments when bundled) is likely a key driving force behind the increased elasticity that composites exhibit (Figs.~\ref{fig:panel1_gt} and ~\ref{fig:panel1_eta}). However, the degree of bundling does not directly correlate with the increased elasticity: while the longest DNA ($ZR \simeq 2.2$) and lowest $\phi_A$ values display the most pronounced peaks in $g_{A,A}$ and $\Pi_{A,A}$ -- indicating the most extreme actin bundling -- it is in fact the intermediate DNA length ($ZR \simeq 1$) and  $\phi_{A} = 0.5$ that displays the strongest increase in elasticity.  

To shed more light on this complex relationship between structure and mechanics, we extract quantitative parameters from the RDF and NCF curves to characterize the composite micro-structure. For instance, the location of the RDF maximum, $r_a$, is a measure of inter-actin spacing in bundles whereas the decay of the NCF curves to $1/2$, $r_b$, measures the lengthscale of alignment or the size of the bundle~\cite{Fitzpatrick2018}. These values, along with the inter-filament spacing for a fully-mixed system, $l_f$, are reported in Table~\ref{tab:ls}. The ratio $r_b/r_a$ is a crude measure of the number of filaments in a bundle, while $r_a/l_f$ indicates the degree to which actin in bundles is packed beyond uniform spacing. As shown, for all DNA lengths, $r_b/r_a$ is maximum when $\phi_A = 0.5$, indicating that the stiffer bundles formed by more filaments play an important role in the composite stiffness (recall that $\phi_A = 0.5$ is where a maximum in stiffness is observed).  For $r_a/l_f > 1$ indicates weak bundling as filaments within bundles are no closer together than when they are uniformly distributed. Conversely, for $r_a/l_f < 1$ actin filaments within bundles are driven closer together than $l_f$, likely forcing connections between bundles to be broken and thus weakening the actin scaffold as a whole. Thus, we conjecture that maximizing composite stiffness requires $r_a \simeq l_f$ , which is achieved for $(ZR, \phi_A) = (\sim 1, 0.5)$.

\begin{table}[h!]
	\begin{tabular}{c | c | c | c | c | c | c }
		$L_{DNA} [\mu$m] & $\phi_A$ & $l_f$ & $r_a$ & $r_b$ & $r_b/r_a$ & $r_a/l_f$ \\
		\hline 
	    & $0.75$ & 4.1$\sigma$ & 6.2$\sigma$ & 13.4$\sigma$ & 2.16 & 1.5 \\
    96  & $0.5$ & 4.7$\sigma$ & 2.5$\sigma$ & 7.5$\sigma$ & 3.0   & 0.53 \\
		& $0.25$ & 5.8$\sigma$ & 3.0$\sigma$ & 4.4 $\sigma$& 1.466 & 0.51 \\
		\hline
        & $0.75$ & 4.1$\sigma$ & 8.2$\sigma$ & 14.2$\sigma$ & 1.73 & 2.0 \\
	39 	& $0.5$ & 4.7$\sigma$ & 4.9$\sigma$ & 16.5 $\sigma$ & 3.36 & 1.04    \\
		& $0.25$ & 5.8$\sigma$ & 3.0$\sigma$ &  4.1$\sigma$& 1.366 & 0.51\\
		\hline
		& $0.75$ & 4.1$\sigma$ & 6.0$\sigma$ & 7.8$\sigma$ & 1.3 & 1.46 \\
	3.67 & $0.5$ & 4.7$\sigma$ & 7.0$\sigma$ & 19.6$\sigma$ & 2.8 & 1.48 \\
		& $0.25$ & 5.8$\sigma$ & 4.6$\sigma$ &  6.9$\sigma$& 1.5 & 0.8\\
		\hline
	\end{tabular}
	\caption{\textbf{Key lengthscales of actin bundling computed from $g_{A,A}(r)$ and $\Pi_{A,A}(r)$}. For each $ZR$ and $\phi_A$ value the following parameters are computed in units of simulation bead size $\sigma$: (i) The average free space between actin beads in a uniformly mixed system $l_f=\rho^{-1/3}$; (ii) the position $r_a$ of the maximum of $g_{A,A}(r)$ for de-mixed states, indicating the inter-filament spacing within bundles; and (iii) the decay length $r_b$ of the nematic parameter $\Pi_{A,A}(r)$ for de-mixed states, indicating the size of bundles. The ratio $r_b/r_a$ is a measure of the number of filaments in a bundle and $r_a/l_f$ measures how closely packed filaments in bundles are compared to filaments in uniformly mixed states.}
	\label{tab:ls}
\end{table}

\dmi{In summary, in this section we have shown that it is not only the balance between actin bundling and connectivity that drives the increase in stress stiffening but also the coupling of the number of entanglements along each chain of flexible and stiff species. Furthermore, we have provided fresh computational evidence of substantial micro-scale dynamics and rearrangements within the composites. These rearrangements are, to the best of our knowledge, not documented in the literature in any composite material and thus prompt the design of new experiments to address them in the future. At present, it is difficult to characterise the macroscopic consequences of these microscale fluctuations because the measurement of stress relaxation in simulations must be averaged over several decades in time and thus encompasses the timescale of structural rearrangement. Thus, we limit ourselves to speculate that there may be a connection between these sudden dynamics and mechanical failure in soft materials. Indeed, it has been observed that peculiar dynamics, such as reversible particle displacements, irreversible rearrangements and heterogeneities often precede network failure~\cite{Aime2018}. Future avenues to characterize the mechanical implications of such rearrangements are discussed below. }

\subsection{Optical Tweezers Microrheology}

\begin{figure*}[t!]
	\centering
	\includegraphics[width=0.9\textwidth]{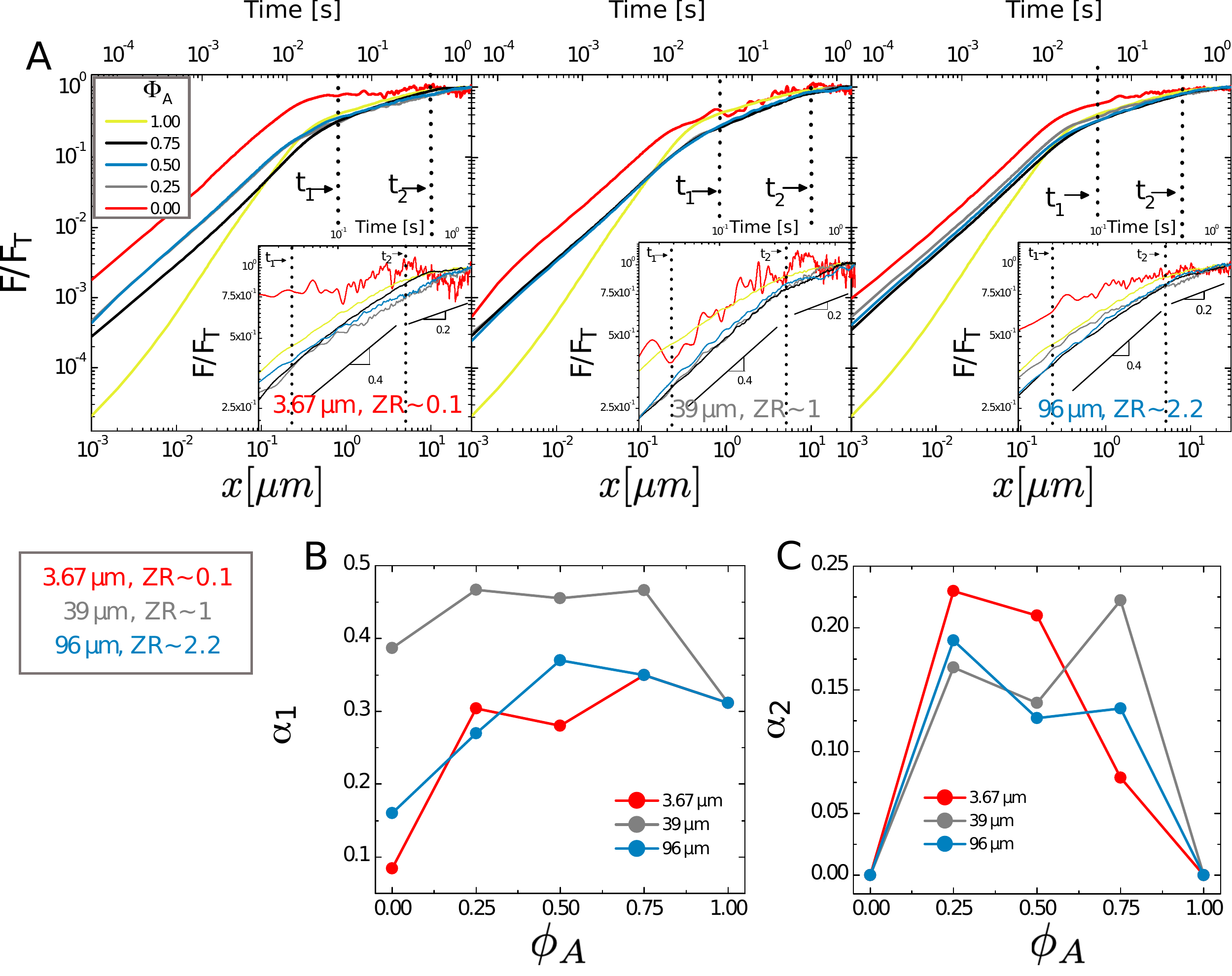}
	\caption{\textbf{Experimentally realized composites exhibit universal sustained elasticity in response to nonlinear forcing.} \textbf{A} Force $F$ as a function of bead displacement $x$ and time $t$, normalized by the terminal value $F_t$, for DNA-actin composites of varying $\phi_A$ (indicated in legend). Each panel displays data for the DNA length indicated. Dashed lines denote times ($t_1$, $t_2$) at which curves crossover to different power-law regimes. Insets: Zoom-in of force near the end of strain. Scale bars show representative scaling exponents for composite systems ($0<\phi_A<1$) .  As shown, upon rescaling $F$ by $F_t$ composites display a universal response with more sustained elasticity (i.e. larger $\alpha_1$ and $\alpha_2$ values) than single-component networks. \textbf{B-C} Dependence of scaling exponents $\alpha_1$ (\textbf{B}) and $\alpha_2$ (\textbf{C}) on $\phi_A$ for different DNA lengths indicated in legends. As shown, composites universally exhibit stiffer, more elastic response (i.e. larger scaling exponents) than single-component networks.}
	\label{fig:panel1_microrheo1}	
\end{figure*}

 To validate our predicted design principles and to extend our understanding of these complex fluids beyond the equilibrium behaviour, we experimentally realize our designed composites and perform nonlinear optical tweezers microrheology measurements (see Materials and Methods).  We choose to perform nonlinear measurements (i.e. large strains and rates) to complement the simulated linear rheology and to test the robustness of actin bundling to strain, as simulations suggest that bundling is relatively weak (few filaments per bundle, only modest demixing). 

In our experiments, a microsphere is optically displaced 30 $\mu$m ($>4\times~L_{actin}$)  through the composite at 20 $\mu$m/s ($\gamma^\cdot \simeq 19.8 s^{-1}\simeq \tau_e^{-1}$) while the force $F$ experienced by the bead during and after the displacement (strain) is recorded (Fig.~\ref{fig:panel1}C). During the strain, force curves for all networks exhibit three distinct regimes with different functional dependence on the bead displacement $x$: an initial steep elastic response until $t_1 \simeq 0.04 s$, a shallower rise $F \sim x^{\alpha_1}$ until $t_2 \simeq 0.5 s$ and a near viscous regime with $F \sim x^{\alpha_2}$ where $\alpha_2 \rightarrow 0$ (see Fig.~\ref{fig:panel1_microrheo1}A). However, by normalising each force curve by its terminal value $F_t$, a clear difference between composites (0 $< \phi_A < $ 1) and pure systems ($\phi_A=0$ or 1) emerges. All composites for each $ZR$ collapse onto a single master curve with systematically larger scaling exponents compared to single-component networks, indicative of a more elastic or stiff response. This non-monotonic dependence of $\alpha_1$ and $\alpha_2$ on $\phi_A$, shown in Fig.~\ref{fig:panel1_microrheo1}B,C, is in broad support of our simulation results, as it demonstrates that composites confer more sustained elasticity compared to pure DNA and actin systems which have lower $\alpha_1$ values and $\alpha_2 = 0$ (i.e. purely viscous response). 

\begin{figure*}[t!]
	\centering
	\includegraphics[width=0.9\textwidth]{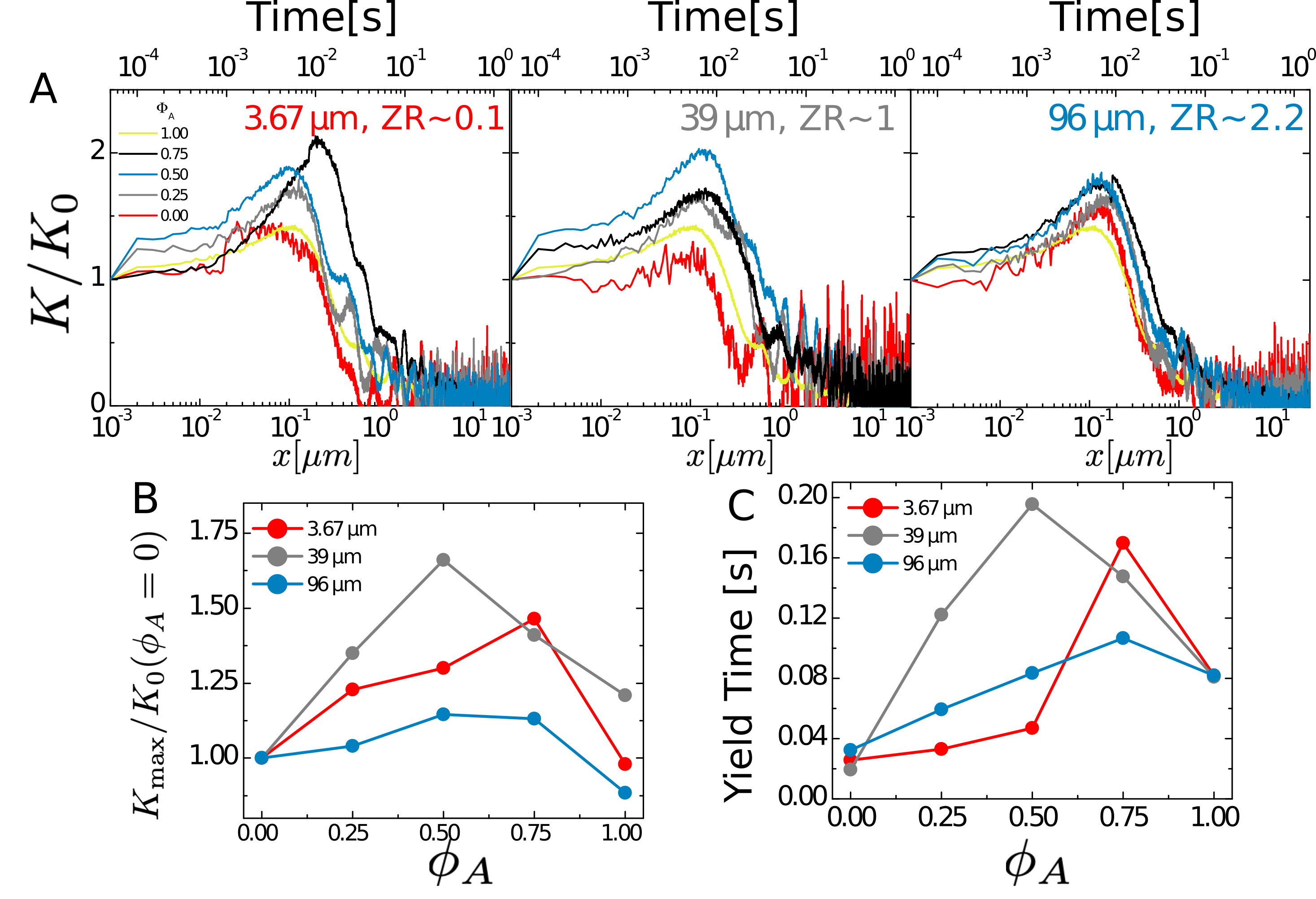}
	\caption{\textbf{The degree of stress-stiffening in composites displays non-monotonic dependence on actin fraction.} \textbf{A} Effective differential modulus $K=dF/dx$, normalized by the initial value $K_0$ for varying $\phi_A$ (indicated in legend). Each panel displays data for the DNA length indicated. \textbf{B} The maximum differential modulus $K_{max}$, normalized by $K_0$, quantifies the degree to which composites stress-stiffen. As shown, stress-stiffening displays a non-monotonic dependence on $\phi_A$ for all DNA lengths (indicated in legend). Normalizing $K_{max}/K_0$ by the corresponding $\phi_A = 0$ value for each DNA length shows that, while the signature non-monotonic dependence is evident for all DNA lengths, the degree of stiffening and the $\phi_A$ value at which maximum stiffening occurs depends on the DNA length.  \textbf{C} The yield time, $t_y$, quantified by the time at which $K = K_0/2e$,  quantifies the time over which composites lose initial elasticity and yield to a viscous regime. Note the non-monotonic dependence on $\phi_A$ and the variation between different DNA lengths is markedly similar to the $K_{max}/K_0(\phi_A=0)$ behavior shown in {C}. Thus, both stress-stiffening and sustained elasticity are similarly tuned by $\phi_A$ and $ZR$ (or $L_{DNA}$).}
   
	\label{fig:panel_microrheo2}	
\end{figure*}

\begin{figure*}[t!]
	\centering
	\includegraphics[width=1\textwidth]{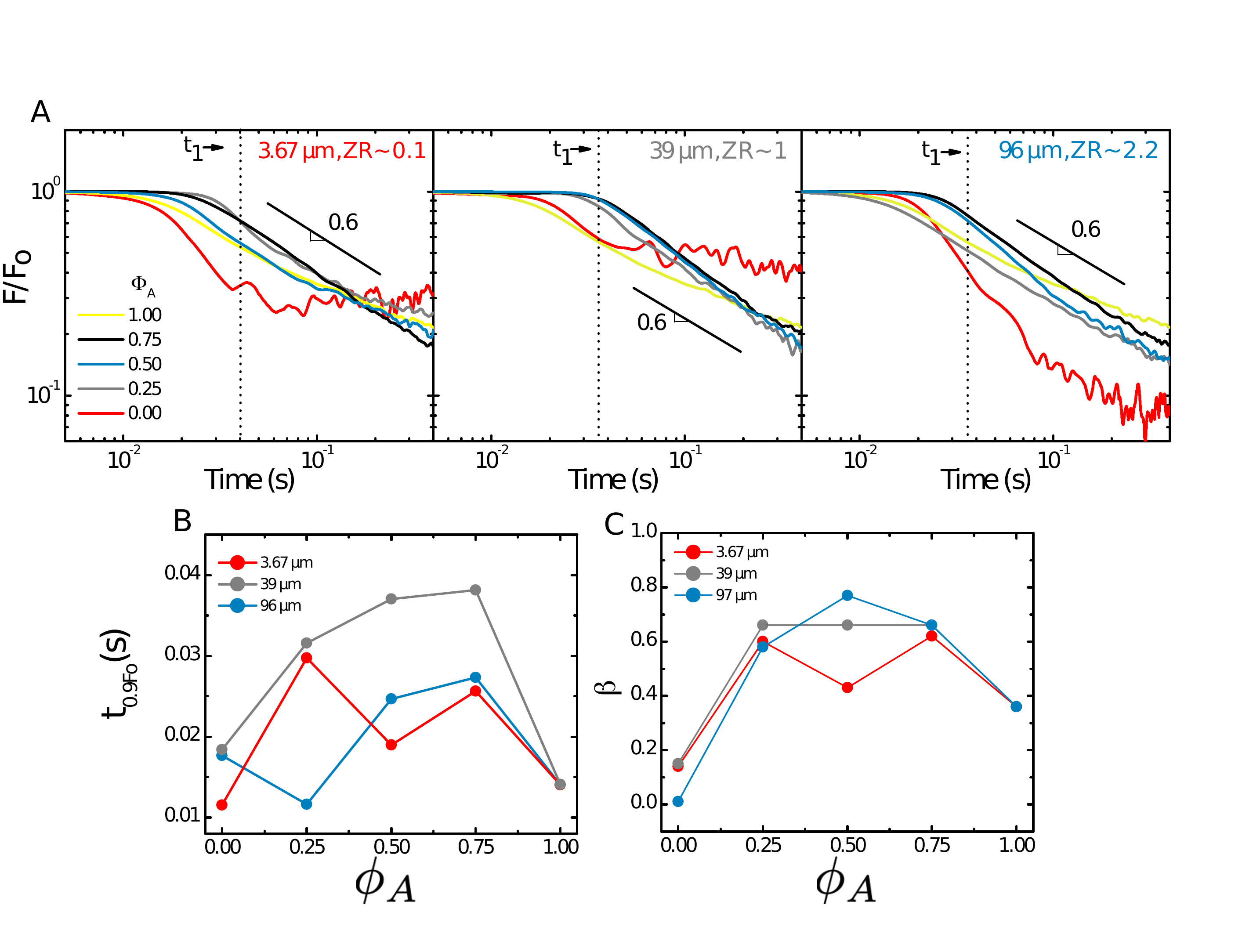}
	\caption{\textbf{Unique force stalling and power-law force relaxation emerges in actin-DNA composites}. \textbf{A}Relaxation of force $F$ as a function of time $t$ following strain, normalized by the corresponding force at $t=0$, $F_0$, for networks of varying $\phi_A$ (see legend). Each panel displays data for the DNA length indicated. Black lines indicate power laws, $F\sim t^{-\beta}$, with exponents listed. All composite networks $(0<\phi_A<1)$ display an initial period of no relaxation (i.e. stalling) until $t_1\simeq 0.04~s $ (dashed line), after which power-law relaxation ensues with $\beta\simeq 0.6$, independent of DNA length. Conversely, single-component networks exhibit near immediate relaxation ($t <0.02~s$), with an initial fast decay until $t_1$ followed by more shallow power-law decays of $\beta_{DNA}\simeq 0.1$ and $\beta_{actin}\simeq 0.4$. \textbf{B} Stalling time, determined as the time at which $F$ drops to $0.9F_0$, as a function of $\phi_A$ for all $ZR$ values. As shown, composites universally display sustained mechano-memory compared to single-component networks. \textbf{C} Scaling exponent $\beta$ as a function of $\phi_A$ for all $ZR$ values, displaying the signature non-monotonic dependence on $\phi_A$. }
	\label{fig:panel_microrheo3}	
\end{figure*}

To further quantify the increased elasticity or stiffness that composites exhibit, and to evaluate the initial response ($t < t_1$), we compute an effective differential modulus $K= dF/dx$. This quantity is a direct measure of the $x$-dependent stiffness or ``spring constant'' of a system. As shown in Fig.~\ref{fig:panel_microrheo2}, in which $K$ is normalized by the corresponding initial value $K_0\equiv K(t=0)$, all systems initially stiffen ($dK/dx > 0$) from a value $K_0$ to a maximum value $K_{max}$, followed by softening ($dK/dx < 0$) and finally yielding ($K \rightarrow 0$). However, the degree of stiffening ($K_{max}/K_0$) as well as the time over which stiffness persists before yielding to more viscous response (i.e. yield time $t_y$) both depend non-monotonically on $\phi_A$, with similar features as seen in the simulated $\eta$ curves (compare Fig.~\ref{fig:panel_microrheo2}B-C with Fig.~\ref{fig:panel1_eta}A-B). Specifically, we find that $K_{max}/K_0(\phi_A = 0)$ and $t_y$ reach global maxima for ($ZR, \phi_A) = $($\sim 1, 0.5$) and another maximum for ($ZR, \phi_A) = $($\sim 0.1, 0.75$). Further, composites comprised of the longest DNA ($ZR \simeq 2.2$) exhibit weaker non-monotonic dependence on $\phi_A$ than for the other two DNA lengths, as seen in simulations (Fig.~\ref{fig:panel1_eta}). Further, while the yield times for single-component systems are close to the predicted entanglement time $\tau_e \simeq 0.04~ s$, the measured $t_y$ for composites is notably higher, suggesting slower relaxation mechanisms and more mechano-memory.

To elucidate how composites relax stress imposed by nonlinear forcing, we measure how the force relaxes following the applied strain, as shown in Fig.~\ref{fig:panel_microrheo3}.  As with the force response during strain, the force relaxations of composites (normalized by the initial value immediately following the strain) are notably distinct from those of single-component systems. Composite systems display force ``stalling'', in which the imposed force does not dissipate, for times up to $t_1 \simeq \tau_e$,  whereas single-component systems begin to relax force nearly immediately following strain ($\sim 0.01~s$). This signature of sustained elasticity (or mechano-memory) is in agreement with simulation results that show distinct elastic plateaus and increased viscosity that are not apparent in single-component systems.  For $t > t_1$, composites display power-law relaxation with scaling exponents $\beta$ that are largely independent of $L_{DNA}$ and $\phi_A$, with an average value of $\beta\simeq0.6~s$. Conversely, DNA and actin systems have weaker scaling exponents of $\sim0.1~s$ and $\sim0.3~s$ respectively. Thus, as shown in Fig.~\ref{fig:panel_microrheo3}C the non-monotonic dependence of mechanics (in this case the scaling exponent $\beta$) on $\phi_A$ is once again preserved.

\dmi{In summary, our experimental results validate our simulations, showing that actin-DNA composites with DNA lengths that span 2 orders of magnitude all exhibit increased elasticity and stiffness compared to single-component systems. They also confirm that the strength of stiffening, as well as the optimum mass composition to achieve maximum stress-stiffening, are tuned by the length of the flexible component. } s 
 
It is notable that comparable micro-strains applied to entangled DNA systems and crosslinked actin networks of similar concentrations have been shown to be sufficient to disentangle DNA~\cite{Chapman2014prl} and break actin crosslinkers~\cite{Gurmessa2016}; on the other hand, here we show that micro-scale bundling in composites is strong enough to withstand such forcing (as evidenced by the increased elasticity and stiffening response). 

\dmi{We further note that we see no signs of spatial or temporal heterogeneities in the force response despite simulations indicating dynamic bundling. In other words, our individual measurements taken at different locations within the sample and at different points in time all display similar force curves. This may be because the bead employed is large compared with the scale of these rearrangements or because the nonlinear forcing disrupts these subtle dynamics.} While outside the scope of the current study, our future work will focus on systematically investigating the dynamic nature of composite structure that simulations predict, and the role that such activity may play in experimentally realized material properties.


\subsection{Theoretical Arguments to Explain Composite Behaviour}

The bundling of actin reported above can be readily explained through theoretical arguments. While in our simulations DNA polymers are effectively flexible ($l_p = 2\sigma \ll L_{DNA}$), actin polymers have a persistence length larger than their contour length ($L_{ACT}$) and diameter ($\sigma$), implying that they behave like nearly rigid rods with large aspect ratio $\alpha=L_{ACT}/\sigma = 280$. 
Attraction of large rigid bodies due to smaller particles or non-adsorbing polymer coils was first discovered by Asakura and Oosawa~\cite{Asakura1954} and then extended to a mixture of polymer coils and rigid rods by Flory~\cite{Flory1978}, who also predicted a change in the isotropic-nematic crossover concentration for such a mixture with respect to the pure solution of rigid rods, which instead follows Onsager's theory~\cite{Tuinier2016}.
Later theoretical and numerical studies further investigated these mixtures through thermodynamic perturbation theory~\cite{Lekkerkerker1994} and Monte Carlo simulations~\cite{Bolhuis1997}.

Within the generalised Asakura-Oosawa framework, polymer coils are assumed to be smaller than the diameter of the rod-like colloids (in our case $\sigma$) and thus excluded from a corona of thickness $R_g$. A key parameter regulating the range of the depletion-attraction is thus $q=R_g/\sigma$ which is generally taken to be $q<1$. However, our simulations and experiments are in the regime $q>1$ for all DNA lengths ($R_g/\sigma \simeq 10$, $33$, $43$), so we should instead view the problem in terms of the (entropic) cost of inserting a thin rod within a coil of size $R_g$. 

Since our actin filaments have a very large aspect ratio ($L_{ACT}/\sigma=280 \gg R_g/\sigma$) we can disregard end-effects and estimate such insertion free energy as the sum of $L_{ACT}/\sigma$ free energies required for inserting a spherical particle of size $\sigma$ within a polymer coil. When the solution of flexible polymers is below the overlap concentration $(c<c^*)$ this can be estimated as $F/(k_BT L_{ACT}) \sim \sigma^{-1} \left(\sigma/R_g\right)^{4/3}$~\cite{Gennes1979a,Sear1997}. Because we are in the limit in which the actin fibres are longer than one DNA coil, then this contribution needs to be multiplied by the volume fraction occupied by DNA in the system, i.e. 
\begin{equation}
\dfrac{F}{k_BT L_{ACT}} \sim \sigma^{1/3} R_g^{5/3} c_{DNA} \, , \label{eq:F_dilute}
\end{equation}
where $c_{DNA}$ is the density of DNA polymers in solution.

In the strongly overlapping regime $(c \gg c^*)$, the relevant length scale is the mesh size $\xi$ which also sets the osmotic pressure $\Pi \sim k_BT/\xi^3$~\cite{Gennes1979a}.  The free energy is then the volume of the particle to be inserted times the osmotic pressure of the entangled network of DNA, i.e. 
\begin{equation}
\dfrac{F}{k_BT L_{ACT}} \sim \dfrac{1}{\sigma} \left( \dfrac{\sigma}{\xi} \right)^3 \sim \sigma^2 c_{DNA}^{9/4} \, ,  \label{eq:F_entangled}  
\end{equation}
where we made use of the fact that $\xi \sim c_{DNA}^{-3/4}$~\cite{Gennes1979a}.

Eqs.~\eqref{eq:F_dilute}-\eqref{eq:F_entangled} tell us that in both regimes the free energy cost of inserting an infinitely long thin rigid rod within a solution of polymer coils scales with the concentration of flexible polymer segments. This is indeed what we observe in Fig.~\ref{fig:entropy_mixing}: bundling increases, i.e. larger $g_{A,A}(r)$ peaks, as $\phi_A$ decreases ($c_{DNA}$ increases) because the free energy cost of actin mixing with DNA becomes increasingly large. 
We also observe that the peaks for the composites with the two longest DNAs are comparable to one another while the shortest DNA exhibits smaller $g_{A,A}(r)$ peaks. We argue that this arises because our two longer DNAs are well within the overlapping regime ($c^*\simeq \, 0.025$ and $0.01$ mg/ml)~\cite{Robertson2006,Laib2006}, and Eq.~\eqref{eq:F_entangled} tells us that in this regime we should indeed see minimal $R_g$ dependence. However, the shortest DNA is in a concentration regime in between those for which Eqs.~\eqref{eq:F_dilute} and \eqref{eq:F_entangled} are valid, with $c^*\simeq 0.13$ mg/ml (or $c/c^* \simeq 1-6$), so we expect actin bundling to depend on $R_g$ as well as $c_{DNA}$. Additionally, in this less entangled regime we expect the $g_{A,A}(r)$ curves to depend on DNA's $R_g$ and therefore $L_{DNA}$, as observed in Fig.~\ref{fig:entropy_mixing}. We argue that, generally speaking, contributions from $c_{DNA}$ (and thus $\phi_A$) as well as $R_g$ (and thus $L_{DNA}$) may play a role in depleting actin filaments from within DNA coils and in turn drive actin bundling in DNA-actin composites.



\section{Discussion and Conclusions}

Polymer composites are formidable complex fluids that are found in both synthetic and natural settings. Deeper understanding of naturally occurring polymer composites as well as the design of futuristic synthetic biomimetic materials require precise knowledge and characterisation of their complex rheological behaviour. In this work we have coupled extensive large-scale Brownian Dynamics simulations with optical tweezers microrheology measurements to investigate the linear and nonlinear mechanical properties of carefully designed blends of DNA and actin polymers. We chose to fix the total mass concentration to a value at which key intrinsic lengthscales of the two species, such as tube radius, are matched. \dmi{This strategy allows us to reduce the dimensionality of the parameter space and to isolate the effects of DNA length and mass composition on the material properties}. \dmi{Importantly, this strategy also allows us to preserve the entanglement length, and in turn keep the number of entanglements per polymer (actin or DNA) constant when varying mass composition.} 

\dmi{One would expect that the mechanical resilience and ability to stress-stiffening would strengthen as the length of the flexible species (i.e. DNA) -- and thus the overall entanglement of each chain --  is increased. Surprisingly, our computational and experimental results rebuff this hypothesis and instead show that maximal stress-stiffening is achieved when the two species are maximally coupled, i.e. have similar number of entanglements per chain. }
\dmi{Our results reinforce the idea that, in composites, the whole is more than the sum of its parts and more is not always better. As such, emergent properties due to non-trivial structural arrangements and physical interactions between species need to be fully characterised in order to drive the discovery and design of the next generation of materials.}


\dmi{To this end, we take advantage of our large-scale BD simulations (run on a supercomputer for the equivalent of about 46 years on a single CPU) to measure precise structural properties of the internal polymer arrangements while monitoring the rheological behaviour of the bulk}. While the stress relaxation and viscosity obtained from the simulations (Fig.~\ref{fig:panel1_gt}) are in broad agreement with experimental microrheology results (Fig.~\ref{fig:panel1_microrheo1}-\ref{fig:panel_microrheo3}), \dmi{our simulations also reveal that the composites undergo unexpected dynamic rearrangements, alternating between partially demixed and mixed states (Fig.~\ref{fig:entropy_mixing})}. 
\dmi{To the best of our knowledge, these rearrangements have not been documented before in any composite system. While their impact on the mechanical properties of our composites cannot be precisely quantified with our current computational and experimental approaches, we expect this intriguing finding to spur future investigations to connect these dynamics to the stochastic behaviour and mechanical failures in soft composite materials~\cite{Aime2018}. }

The partially demixed state is characterised by bundling of actin fibres which can be explained by simple theoretical arguments and generalisations of the Asakura-Oosawa model for depletion-induced attraction. 
This phenomenon is entropically driven by DNA and leads to stiffening of the actin network which acts as a scaffold for the more flexible DNA network. 
However, stronger actin bundling does not simply map to a stiffer mechanical response, as seen in the different behaviours of bundling and elasticity as a function of $ZR$ and $\phi_{A}$ (Figs.~\ref{fig:panel1_gt} and \ref{fig:entropy_mixing}). In particular, while the longest DNA and lowest actin fraction lead to the most pronounced bundling, it is intermediate values of both parameters that confer the most elastic response. 


Within the actin-devoid gaps in the composites (which are larger with more bundling), only DNA can contribute to the mechanical response, so the degree of DNA entanglements (i.e. $Z_{DNA}$) must then play a major role in the mechanical response.
In particular, the longest DNA molecules considered here ($Z_{DNA} \simeq 22$) are well-entangled and therefore can provide structural stability to the material. For this reason the non-monotonic dependence of mechanics on $\phi_A$ is weakest for this DNA length. Conversely, the shortest DNA ($Z_{DNA} \simeq 1$) is not entangled enough to provide rigidity without an actin scaffold, so the non-monotonic dependence peaks at higher $\phi_A$ values than the longer DNA. 
It is only for the intermediate length of DNA ($Z_{DNA} \simeq 10)$, in which the number of entanglements per chain for DNA and actin are closely matched ($ZR = Z_{DNA}/Z_{actin} \simeq 1$), that there are sufficient DNA entanglements to provide some structural support in regions void of actin; yet, they are still relatively weak such that entanglements between DNA and actin fibers are necessary for enhanced stiffness and elasticity. 



\dmi{In summary, our collective approach and results elucidate important and a priori unexpected physical principles for the design of composites made of flexible and rigid polymers. Intriguingly, these systems are found to be tunable yet robust at the same time. They are robust because they display a non-monotonic dependence of the mechanical response on the mass composition that is broadly preserved across the entire parameter space probed. They are tunable because the details, such as the amplitude of the non-monotonic response, degree of actin bundling, and mass composition at which maximal elasticity is achieved, can be altered by varying the length of the flexible species. } 

\dmi{The most important and surprising finding of this work is that composites in which the two polymer species are maximally coupled, rather than maximally entangled, yield the most pronounced stress stiffening and elasticity.}
Our results -- which reveal previously unappreciated physical phenomena at play in composite systems such as micro-scale phase separation, dynamical rearrangements and competition between bundling and pervasiveness -- further elucidate the interactions between distinct polymer species in naturally occurring and biomimetic polymer composites and pave the way for the systematic design of next-generation multi-functional composite materials.


\section{Materials and Methods}
\subsection{Simulations}
\textbf{A. Computational Details} 
DNA and actin filaments are simulated as coarse-grained Kremer-Grest bead-spring polymers~\cite{Kremer1990,Michieletto2019} where each bead represents $\sigma=25$ nm. Excluded volume between beads is accounted for a shifted-and-truncated Lennard-Jones potential
\begin{equation}\label{eq:LJ}
	U_{\mathrm{LJ}}(r) = \left\{
	\begin{array}{lr}
		4 \epsilon \left[ \left(\frac{\sigma}{r}\right)^{12} - \left(\frac{\sigma}{r}\right)^6 + \frac14 \right] & \, r \le r_c \\
		0 & \, r > r_c
	\end{array} \right. \, ,
\end{equation}
where $r$ denotes the separation between the bead centers.
The cutoff distance $r_c=2^{1/6}\sigma$ is chosen so that only the repulsion part of the Lennard-Jones is used. The energy scale is set by $\epsilon=k_BT$.
Consecutive monomers along the polymers are connected through finitely extensible nonlinear elastic (FENE) springs:
\begin{equation}\label{eq:Ufene}
	U_{\mathrm{FENE}}(r) = \left\{
	\begin{array}{lcl}
		-0.5kR_0^2 \ln\left(1-(r / R_0)^2\right) & \ r\le R_0 \\ \infty & \
		r> R_0 &
	\end{array} \right. \, ,
\end{equation}
where $k = 30\epsilon/\sigma^2$ is the spring constant and $R_{0}=1.5\sigma$ is the maximum extension of the FENE bond.
To model the different rigidity of DNA and actin, we introduce an additional bending energy penalty between consecutive triplets of neighbouring beads along the chain:
\begin{equation}\label{eq:Ubend}
	U_{\mathrm{bend}}(\theta) = \dfrac{k_B T l_p}{\sigma} \left(1+ \cos \theta \strut\right) \, ,
\end{equation}
where $\theta$ is the angle formed by consecutive bonds. The persistence length is $l_p ({\rm DNA})= 2 \sigma= 50$ nm for DNA and $l_p({\rm actin})=400 \sigma=10 \, \mu$m for actin.

The motion of each bead is evolved through a Langevin equation 
\begin{equation}
    m \dfrac{d^2 \bm{r}}{dt^2} = -\zeta \dfrac{d \bm{r}}{dt} - \nabla U + \sqrt{2 k_BT \zeta} \bm{f}
\end{equation}
with friction $\zeta= 3 \pi \eta \sigma$, Gaussian white noise $\bm{f}$ satisfying the fluctuation-dissipation theorem and temperature $T$ in LAMMPS~\cite{Plimpton1995}. Using the viscosity of water $\eta=1 cP$ the typical diffusion time of a bead is $\tau_B =\sigma^2/D = 3\pi\eta\sigma^3/k_BT \simeq 0.04$ ms and the total simulation runtime can reach up to $10^7 \tau_B \simeq $ 400 seconds. 

\textbf{B. System Parameters}
We consider DNA molecules with contour lengths $L_{DNA} = 3.67, \, 39, \, 96 \, \mu \rm{m}$ 
and actin polymers with contour length $L({\rm actin})=7\, \mu$m 
The total concentration is set to $c=0.8$ mg/ml and we vary the relative fraction of actin $\phi_A = c_{actin}/c = 0, \,0.25, \,0.5, \,0.75 , \, 1$. 
All of the systems with $\phi_A=1$ are independent realisations of the same system of entangled actin.
It should be noted that each polymer bead ($\sigma=25$ nm) represents either 75 DNA base-pairs or 9.2 actin monomers and that the molecular weight of one DNA bp is 650 g/mol whereas that one of an actin monomers is 42000 g/mol. This implies that a different mass is effectively represented by one DNA bead and one actin bead in our simulations: more specifically each actin bead weighs about 8 times a DNA bead ($= (9.2 \times 42 \,kDa) /(75 \times 650\, Da)$). Yet, in our simulations the dynamics is overdamped after few integration time-steps so we do not expect the mass (inertia) of the beads to have an effect on the dynamics in either simulations or the real systems.  The simulations are performed in boxes of fixed size $L = 100 \sigma$ = 2.5 $\mu$m, and because of the different in bead mass and fixed mass concentration there is a variation in the total number of beads: the 100\% DNA solution has $\sim$153,300 beads whereas the system with 100\% actin has $\sim$19'040 beads. The mixed cases have intermediate values. 


For the production runs (after equilibration), we run 3 DNA lengths and 5 conditions for $\sim$5 weeks on a 32 CPU computer, totalling at $\sim$0.5M CPU-h or 46 years on a single CPU machine.  

\textbf{C. Equilibration}
The polymers are initialised as perfectly flexible random walks within a box of size $L=100 \sigma= 2.5 \mu$m. Each system is then equilibrated, i.e. a full MD simulation is performed until the mean square displacement (MSD) of the centre of mass of the polymers have travelled at least once their typical size, measured as the gyration radius $R_g$. Also, $R_g$ is monitored and checked to have reached steady state before starting the production run.
\dmi{By inspecting the conformations of actin filaments at equilibrium we have also checked that they do not frequently interact with their own tail albeit being longer ($7 \mu$m) than the simulation box (2.5 $\mu$m). This is because they are not treated as rigid rods but as semi-flexible chains thus allowing for some (small) bending. It should be noted that the calculation of the radial distribution function (RDF) and nematic correlation function (NCF) purposely avoid accounting for beads belonging to the same chain and thus the peaks in RDF and NCF are genuinely formed by different chains that come together in bundles.
}

\subsection{Experiments}
\textbf{A. Proteins, DNA, and microspheres}: Monomeric rabbit skeletal muscle actin (Cytoskeleton, AKL99) was stored at $-80^oC$ in G-buffer [2 mM Tris pH 8.0, 0.2 mM ATP, 0.5 mM DTT, 0.1 mM CaCl$_2$]. Linear double-stranded DNA of lengths 11 kbp (3.67 $\mu$m), 115 kbp (39 $\mu$m) and 289 kbp (96 $\mu$m) were prepared via replication of supercoiled plasmids (11 kbp) and bacterial artificial chromosomes (BACs, 115, 289 kbp) in Escherichia coli, followed by extraction, purification and enzymatic linearization as described previously [9]. BamHI and Mlu1 (New England Biolabs) were used to linearize the plasmid  and BACs respectively. Following linearization, DNA was dialyzed into G-buffer and stored at $4^oC$. Carboxylated polystyrene microspheres (Polysciences), of diameter $d$ = 4.5 $\mu$m, were coated with Alexa-488-BSA, as previously described [10], to inhibit binding interactions with the polymers and visualize beads during measurements. The capping protein gelsolin, which was used to control the actin filament length, was stored at $-80^oC$ in [10 mM Tris pH 7.5, 10 mM NaCl, 0.1 mM MgCl2, 1$\%$ (w/v) sucrose, 0.1$\%$ (w/v) dextran].

\textbf{B. Sample Preparation}: For experiments, actin monomers, DNA, and a trace amount of microspheres were mixed in P-buffer [10 mM Tris pH 7.5, 50 mM KCl, 2 mM MgCl$_2$, 1 mM ATP] for a final polymer concentration of 0.8 mg/ml.  \dmi{P-buffer provides good solvent conditions such that DNA assumes swollen random coil configurations~\cite{Robertson2006}. Semiflexible actin filaments assume extended configurations.} Gelsolin was added at a concentration needed to cap actin filaments to $L$ = 7 $\mu$m, using the relationship $L = (330 R_{GA})^{-1}$ where $R_{GA}$ is the molar ratio of gelsolin to actin [11]. This mixture was pipetted into a 100-$\mu$m thick sample chamber comprised of a microscope slide, $\sim$100 $\mu$m layer of double-sided tape, and a glass coverslip to accommodate a 20 $\mu$L sample. The chamber was sealed with epoxy and the sample was allowed to polymerize and equilibrate for 30 min before measurement. \dmi{Following the 30 min equilibration period, the composites displayed no signs of aging over a 24 hour period. }

\dmi{We note that the presented data is for composites in which actin monomers were polymerized in the sample chamber in the presence of DNA. As crowding effects of the DNA could conceivably affect the polymerization efficiency and final length of the actin filaments, we also performed control experiments in which actin was polymerized and capped prior to mixing with the DNA and loading into the sample chamber. There were no statistically significant differences between results from both preparations. Because DNA and actin are both biological polymers with stability that is temperature-dependent, we did not investigate the temperature dependence of our results. Temperatures above $37^\circ$C can compromise the stability of the polymers and temperatures below room temperature can inhibit actin polymerization. However, within the usable temperature range, we do not expect the dependence of the mechanical response on DNA length and actin fraction to change.}

\dmi{Finally, we note that both actin and DNA are polyelectrolytes so their interactions could be both entropic and enthalpic. However, based on our previous work on entangled single-component solutions of DNA and actin~\cite{Robertson2006,Chapman2014prl,Falzone2015,Gurmessa2016} we assume the interactions to be largely entropic. Specifically, we have shown that similar ionic conditions are sufficient to screen the negative charge of DNA such that it can be treated as a neutral polymer with a salt-dependent effective diameter~\cite{Robertson2006}. Further, both single-component systems have been shown to display dynamics that are in indicative of entangled neutral polymers with no signs of charge effects. Finally, we have also shown that the diffusion and conformations of DNA molecules under crowding conditions are entropically rather than enthalpically driven~\cite{Chapman2015,Gorczyca2015,Mardoum2018}. As such, we model both DNA and actin as neutral polymers with purely steric interactions.}

\textbf{C. Microrheology Measurements}: We applied microscale strains to the composites by using an optical trap and a piezoelectric nanopositioning stage to drag a microsphere through the composite (see Fig.~\ref{fig:panel1}B). The custom-built force-measuring trap was formed by a 1064 nm Nd:YAG fiber laser focused with a 60x 1.4 NA objective. The optical trap construction and its calibration for precision force measurements and bead displacements have been reported previously [3,12]. During each measurement, a microsphere embedded in the composite was displaced 30 $\mu$m at a speed of $20~\mu$m/s. The resulting force the composite exerted on the trapped bead was measured before (5 s), during (1.5 s) and after (20 s) the bead displacement using a position sensing detector to measure the exiting laser deflection, which is proportional to the force [3] (Fig. ~\ref{fig:panel1}B).  Laser deflection and stage position were recorded at a rate of 20 kHz. Custom-written LabView and MATLAB scripts were used to execute the experiments, collect force and stage position data, and analyze the data. For each composite, 20 different measurements were carried out using different microspheres in different regions of the composite. All data presented are ensemble averages of the individual measurements for each composite. 

%

\subsection{Acknowledgements}
The authors would like to thank Richard Sear for thoughtful comments and interesting discussions. This research was funded by an AFOSR Biomaterials Award (No. FA9550-17-1-0249) and an NSF CAREER Award (No. 1255446) awarded to RMR-A.

\subsection{Author Contributions}
RMR-A conceived the idea and supervised the experiments. RF performed experiments. DM designed and performed simulations and data analysis. RMR-A and DM interpreted the data and wrote the paper. 

\bibliographystyle{nar}
\bibliography{library}

\end{document}